\documentstyle[preprint,eqsecnum,aps]{revtex}  

\begin{document}

\draft

\title{Path Integrals and Pseudoclassical Description for
Spinning Particles in Arbitrary Dimensions}

\author{Dmitri M. Gitman\thanks{e-mail: gitman@snfma1.if.usp.br; 
fax: 55-11-8186833; tel: 55-11-8134257}}

\address{Instituto de F\'{\i}sica, Universidade de S\~ao Paulo \\ 
Caixa Postal 66318-CEP 05389-970-S\~ao Paulo, S.P., Brasil
}

\date{\today}

\maketitle

\begin{abstract}

\end{abstract}
The propagator of a spinning particle in external Abelian field and 
in arbitrary dimensions is presented by means of a path integral. The
problem has different solutions in even and odd dimensions. In even
dimensions the representation is just a generalization of one in four
dimensions (it has been known before). In this case a gauge invariant part of the 
effective action in the  path integral has a form of the standard (Berezin-Marinov) 
pseudoclassical action. In odd  
dimensions the solution is presented for the first time and, in
particular, it turns out that the gauge
invariant  part of the effective action  differs from the standard
one. We propose this new action as a candidate to describe
spinning particles in odd dimensions. Studying the hamiltonization of
the pseudoclassical theory with this action, we show that  the
operator quantization leads to adequate minimal quantum theory of spinning particles
in odd dimensions.  In contrast with the models proposed formerly in this case  
 the new one admits both the  operator and the path integral
quantization. Finally the consideration is generalized for the case of
the particle with anomalous magnetic moment.

\section{Introduction}
Construction of classical and pseudoclassical models of relativistic
particles as well as their quantization (first quantization) attracts
attention already for a long time due to various reasons. One can
mention here both a natural interest to fill in gaps in fundamental
theoretical constructions  and well-known close relation to the
string theory, where the first quantization remains until now the main
way to quantum description. It seams that in four dimensions the work
in this direction is almost done, namely, there exist wide-acceptable
classical and pseudoclassical models (actions) for particles of different
spins. Operator quantization of the models lead to the quantum
mechanics which are equivalent to one-particle sectors of quantum
field theories, whereas the path integral quantization reproduces the
corresponding propagators. The main problem in such  models is
connected with spins description. Usually they introduce Grassmann
variables to describe spins, that is why the models are called
pseudoclassical. The basic pseudoclassical model for all those
constructions in four dimensions is one for the spinning (spin
one-half) particle, proposed first by Berezin and Marinov (BM) \cite{BM}, and
then discussed and studied in numerous papers \cite{All1}. It was
shown \cite{All2} that the Dirac propagator in four dimensions can be
presented as a path integral (with appropriate gauge fixing
conditions) of $\exp (iS)$, where the effective action  $S$ is BM action.  

If one extends formally BM model to arbitrary space-time 
dimensions, one can discover that in even dimensions it serves perfectly
to describe particles with spin one-half, whereas in odd dimensions
difficulties appear. Technically they are connected with the absence of
the analog of $\gamma^5$-matrix in odd dimensions. Indeed, in the BM
action in $D$ dimensions there are $D+1$ dynamical Grassmann variables (see
(\ref{c3}))  (a Lorentz vector $\psi^{\mu},\;\mu=0,1,\ldots,D-1$, and a
Lorentz scalar $\psi^D$)  to describe spinning degrees of freedom. The
corresponding operators $\hat{\psi}^\mu,\,\hat{\psi}^D$ obey the
Clifford algebra 
\begin{equation}\label{1}
[\hat{\psi}^k,\hat{\psi}^n]_+=
-\frac{1}{2}\eta^{kn},\;k,n=0,1,\ldots,D\,.
\end{equation}
In even dimensions $D=2d$ one can always realize (\ref{1}) by the
choice:
$\hat{\psi}^\mu=\frac{i}{2}\gamma^\mu,\;\hat{\psi}^D=\frac{i}{2}\gamma^{D+1}$,
where $\gamma^\mu$ are $2^d\times 2^d$ $\gamma$-matrices in $D=2d$
dimensions and  $\gamma^{D+1}$ is an analog of $\gamma^5$ in four dimensions
(see (\ref{b2})). Thus, one gets the minimal quantum theory for
spinning particle with $2^d$ component Dirac equation. In $D=2d+1$
 the above mentioned possibility does not exist (there is no
$\gamma^{D+1}$ there) and one has
to realize (\ref{1}) by means of $\gamma$-matrices from the higher
dimensions, for example, by means of $2^{d+1}\times 2^{d+1}$
$\gamma$-matrices in $D+1$ dimensions. Thus, one obtains a pair of
Dirac equations, describing particles with spins $1/2$ and $-1/2$. In
odd dimensions these are different particles, related to different
irreducible representations of the Lorentz group. Thus, the BM model
in odd dimensions does not reproduces the minimal quantum theory in
course of the operator quantization.  One
meets the same kind of difficulties when constructing a path integral representation for
spinning particle propagator, generalizing, for example, the approach of
the paper \cite{FG} to  odd dimensions.  All that indicates  that
 BM action has to be modified  in odd
dimensions in order to provide both minimal operator and path integral
quantization. One has to remark here that besides of  general
motivations mentioned above, consideration of the odd-dimensional case
is important in relation with the corresponding field theory
(especially in $2+1$ dimensions). In recent  years the latter theory  attracts great
attention due to nontrivial topological properties and the possibility
of  existence of particles with fractional spins and exotic
statistics (anyons).

Recently, there were proposed three different types of pseudoclassical
actions to describe the massive spinning particles in odd-dimensional
space-time, two in  \cite{Pl1,Pl2} respectively and the third one in
\cite{GGT1}. The first one is classically equivalent to BM action,
extended to odd dimensions. It is P- and T- invariant on the classical
level and the violation of the symmetry takes place only on quantum
level, so that an anomaly is presented. No path integral was written 
with this action. Moreover, as was remarked by the authors in
\cite{Pl2}, the P- and T- invariance of this action makes it difficult
to understand how the path integral approach can take care of the
difference between spins $1/2$, and $-1/2$. Another action \cite{Pl2}
is already P- and T- noninvariant and reproduces the adequate quantum
theory in course of  quantization. The authors introduced two
additional (to those which describe spin) 
dynamical Grassmann variables trying to avoid well-known
difficulties  in direct classical treatment of the model (see discussion in
\cite{Pl3,GGT2}). However the action is not supersymmetric. 
 In the papers \cite{GGT1} a different action was
proposed, which is a natural dimensional reduction from the
even-dimensional massless (Weyl particles \cite{GGT3,GG} in even
dimensions) case.  The action is supersymmetric, P- and T-
noninvariant and can be extended to describe higher spins in odd
dimensions \cite{GT4}. However, as in two previous cases no path
integral quantization of the model was given.

In the present paper we have succeeded to construct the path integral
representation for spinning particle propagator both in arbitrary even-
and odd-dimensional cases. In even dimensions we followed the approach
of our paper \cite{FG}, where the path integral in four dimensions was
constructed, using a super-generalization of the Schwinger
proper-time representation \cite{Sch} for the inverse operator, in which 
the proper time is presented by  a pair of even and odd variables. In
this case the effective action, which we extract from the path
integral, coincides with BM one, extended to arbitrary
even-dimensional case. In odd dimensions we have used a different
technical trick to write the path integral representation. Namely, in
this case one has to use a more complicated super-generalization of
the Schwinger representation, where the
proper-time has already one even and two Grassmann components. Thus, for the
first time, we get  a path integral representation for the spinning particle
propagator in odd dimensions. Extracting a gauge-invariant part of the
effective action from the
path integral, we get a new  pseudoclassical action to describe
spinning particles 
in odd dimensions. Since the path integral quantization is already
done by the construction, we have only to verify that the operator quantization, being
applied to the new action,  leads to the minimal quantum theory . To
this end we analyse  the Hamiltonian structure of the theory,
deriving all the constraints and Dirac brackets. Then we present an
explicit realization of the  quantization procedure to get the Dirac
equation in odd dimensions. In the end we discuss the peculiarities of the
new representations obtained. Finally the consideration is generalized for the case of
the particle with anomalous magnetic moment.

\section{Path integral representation for relativistic particle propagator
in arbitrary dimensions}

\subsection{Scalar case in $D$ dimensions}

Let us first discuss briefly  path integral representation for scalar particle
propagator in order to make then more transparent for the reader the problems
connected with spinning degrees of freedom in arbitrary
dimensions.  We consider the particle placed into arbitrary external electromagnetic
field with potentials $A_{\mu}$, that makes the problem non-trivial
and allows one to reveal the features connected with electromagnetic
nature of the particle. As it is known, the propagator of the scalar particle,
interacting with an external electromagnetic field, is the causal
Green function $D^c(x,y)$ of the Klein-Gordon equation,
\begin{equation}\label{b} 
\left( \hat{{\cal P}}^2 - m^2\right)
D^{c}(x,y)=-\delta^D(x-y),
\end{equation}
where $\hat{{\cal P}}_{\mu}=i\partial _{\mu}-gA_{\mu}(x)$, 
$\mu=0,\ldots,D-1$, and the Minkowski tensor is $\eta_{\mu\nu}=
{\rm diag}(\underbrace{1,-1,\ldots,-1}_{D})$. Following Schwinger
\cite{Sch}, one can present $D^c(x,y)$ as a matrix element of an
operator $\hat{D}^c$,
\begin{equation}\label{1b}
D^c(x,y)=<x|\hat{D}^c|y>,
\end{equation}
where $|x>$ are eigenvectors for some self-conjugated operators of
coordinates $X^{\mu}$; the corresponding canonical-conjugated
operators of momenta are $P_{\mu}$, so that: 
\begin{eqnarray}\label{2b}
&X^\mu|x> = x^\mu |x>\,, \,\,\,\,\,\,\,\, 
 <x | y > = \delta^D(x-y)\,, \,\,\,\,\,\,\,\, 
\int|x><x|dx = I\,,&   \nonumber \\
& \left[P_\mu,X^\nu \right]_- = - i \delta_\mu^\nu\,, \,\,\,\,\,\,\,\,
P_\mu|p> = p_\mu |p> \,, \,\,\,\,\,\,\,\, 
<p | p' > = \delta^D(p-p')\,,&  \nonumber \\
&\int|p>< p|dp = I\,, \,\,\,\,\,\,\,\,
<x |P_\mu| y > = -i\partial_\mu\delta^D(x-y)\,,\,\,\,\,\,\,\,\,
<x | p > = \frac{1}{(2\pi)^{D/2}}e^{ipx}\,,&  \nonumber \\ 
&\left[\Pi_\mu,\Pi_\nu \right]_- = - igF_{\mu\nu}(X)\,, \,\,\,\,\,\,\,\, 
\Pi_\mu = -P_\mu - g A_\mu(X)\,.& 
\end{eqnarray}
Equation (\ref{b}) implies
$\hat{D}^c=\hat{F}^{-1}\,,\;\;\hat{F}=m^2-\Pi^2\;.$ 
Now one can use the Schwinger proper-time representation for the
inverse operator
\begin{equation}\label{3b}
\hat{F}^{-1}=i\int_{0}^{\infty}e^{-i\lambda (\hat{F}-i\epsilon)}\,d\lambda\;,
\end{equation}
where $\lambda$ is the proper-time and the infinitesimal quantity
$\epsilon$ has to be put to zero at the end of  calculations.
Thus, we get for the Green function (\ref{1b}) 
\begin{eqnarray}\label{5b}
&&D^{c}=D^{c}(x_{out},x_{in}) =i\int_0^\infty \,  
\langle x_{\rm out} | e^{-i\hat{\cal H}(\lambda)}|x_{\rm in} 
\rangle \,d\lambda\,, \\
&&\hat{\cal H}(\lambda)= \lambda \left(m^2-\Pi^2\right)\,. \nonumber
\end{eqnarray}
Here and in what follows we include the factor 
$-i\epsilon$ in $m^2$. Now one can present the matrix element entering in the expression
(\ref{5b}) by means of a path integral.  So, as usual, we write 
$\exp (-i\hat{{\cal H}}) = \left[\exp (-i\hat{{\cal H}}/N) \right]^N $ 
and then insert $(N-1)$
resolutions of identity $\int|x><x|dx = I$ between all the 
operators $\exp (-i\hat{{\cal H}}/N)$. Besides, we introduce
$N$ additional integrations over $\lambda$  to transform then the
ordinary integrals over these variables into the corresponding path-integrals,
\begin{eqnarray}\label{6b}
&& D^{c} = i \lim_{N\rightarrow \infty}\int_0^\infty d\, \lambda_0 
\int_{-\infty}^{+\infty} d\, x_1 ... d\, x_{N-1}
d\, \lambda_1 ... d\, \lambda_N  \nonumber \\
&&\times \prod_{k=1}^{N}\langle x_k |
e^{-i\hat{{\cal H}}(\lambda_k)\Delta \tau} | x_{k-1} \rangle 
\delta(\lambda_k-\lambda_{k-1})\; ,
\end{eqnarray}
where $\Delta \tau = 1/N$, $x_0=x_{\rm in}$, $x_N=x_{\rm out}$. Bearing in 
mind the limiting process, one can calculate the matrix elements 
from (\ref{6b}) approximately,
\begin{equation}\label{7b}
\langle x_k |
e^{-i\hat{{\cal H}}(\lambda_k)\Delta \tau} | x_{k-1} \rangle \approx
\langle x_k |
1{-i\hat{{\cal H}}(\lambda_k)\Delta \tau} | x_{k-1} \rangle,
\end{equation}
using the resolution of identity $\int|p><p| d\,p=I$. In this 
connection it is  important to notice that the operator
$\hat{{\cal H}}(\lambda)$ has originally the symmetric form in the 
operators $X$ and $P$. Indeed, the only one term in 
$\hat{{\cal H}}(\lambda)$, which contains products of these operators
is $[P_\alpha,A^\alpha(X)]_+$. One can verify that this is
maximal symmetrized expression, which can be combined from entering operators
(see remark in  \cite{DWi}). Thus, one can write
$\hat{{\cal H}}(\lambda) = {\rm Sym}_{(X,P)}\,\, 
{\cal H}(\lambda,X,P)$,
where ${\cal H}(\lambda,x,p)$ is the Weyl symbol of the 
operator $\hat{{\cal H}}(\lambda)$,
\[
{\cal H}(\lambda,x,p) =  \lambda \left(m^2-{\cal P}^2\right)\,,\;\;
{\cal P}_\mu= -p_\mu-gA_\mu(x)\;.
\]
That is a general statement \cite{Ber1}, which can be easily checked in that
concrete case by direct calculations, that the matrix elements (\ref{7b})
are expressed in terms of the Weyl symbols in the middle point 
$\overline{x}_k = (x_k+x_{k-1})/2$.
Taking all that into account, one can see that in the limiting process the 
matrix elements (\ref{7b}) can be replaced by the expressions 
\begin{equation}\label{8b}
\int \frac{d\,p_k}{(2\pi)^D}\exp\left\{ i \left[
p_k\frac{x_k-x_{k-1}}{\Delta\tau} - {\cal H}(\lambda_k,
\overline{x}_k,p_k) \right]\Delta\tau \right\}\,.
\end{equation}
Using the integral representation for the $\delta$-functions, we get
for the right side of (\ref{6b})
\begin{eqnarray}\label{9b}
&& D^{c} = i \lim_{N\rightarrow \infty}\int_0^\infty d\, \lambda_0 
\int_{-\infty}^{+\infty} d\, x_1 ... d\,
x_{N-1}\frac{d\,p_1}{(2\pi)^D}\cdots
\frac{d\,p_N}{(2\pi)^D}
d\, \lambda_1 ... d\, \lambda_N  \nonumber \\
&&\times \frac{d\,\pi_1}{(2\pi)}\cdots \frac{d\,\pi_N}{(2\pi)}
\exp\left\{i\sum_{k=1}^{N}\left[p_k\frac{ x_k-x_{k-1}}{\Delta
\tau}-{\cal H}(\lambda_k,\overline{x}_k,p_k)+\pi_k\frac{ \lambda_k-
\lambda_{k-1}}{\Delta
\tau}\right]\Delta \tau\right\}\;.
\end{eqnarray}
The above  expression  is, in fact, the definition of the Hamiltonian
path integral for the scalar particle propagator,
\begin{equation}\label{10b}
D^c=i\int_{0}^{\infty } d\,\lambda_0\int_{x_{in}}^{x_{out}}D\,x\int_
{\lambda_0}D\,\lambda\,
 \int D\,p\,D\,\pi\,\exp\left\{i \int_{0}^{1} [\lambda ({\cal
P}^2-m^2)+p\dot{x}+\pi\dot{\lambda}]d\,\tau   \right\}\;,
\end{equation}
where ${\cal P}_{\mu}=-p_{\mu}-gA_{\mu}(x),\;\dot{x}=\frac{d}{d\tau}x$,
and so on. The functional
integration goes over the trajectories
$x^{\mu}(\tau),p_{\mu}(\tau),\lambda(\tau),\pi (\tau)$, parameterized
by some invariant parameter $\tau\in[0,1]$ and obeying the boundary
conditions $x(0)=x_{in},\;x(1)=x_{out},\;\lambda (0)=\lambda _0$. 
To go over to the Lagrangian form of the path integral, one has to
perform the integration over the momenta $p$. In fact, the result can  be
achieved by means of the replacement,
$p_{\mu} \rightarrow
-p_{\mu}-(\dot{x}_{\mu}/2\lambda)-gA_{\mu},\;\;e=2\lambda $. Thus, we get
\begin{eqnarray}\label{11b}
&&D^c=\frac{i}{2}\int_{0}^{\infty}d\,e_0\int_{x_{in}}^{x_{out}}D\,x\int_{e_0}
M(e) D\,e \int D\,\pi \nonumber \\
&&\times \exp\left\{i \int_{0}^{1}\left [
-\frac{\dot{x}^2}{2e}-e\frac{m^2}{2}
-g\hat{x}_{\mu}A^{\mu}+\pi \dot{e}\right]d\,\tau   \right\}\;,
\end{eqnarray}
where the boundary conditions
$x(0)=x_{in},\;x(1)=x_{out},\;e(0)=e_0$ 
are supposed and the measure $M(e)$ has the form   
\begin{equation}\label{12b}
M(e)=\int D\,p\exp\left\{ \frac{i}{2}\int_{0}^{1} ep^{2}d\tau\right\} \;.
\end{equation}
A discussion of the role of the measure (\ref{12b}) one can find in \cite{FG}.

\subsection{Spinor case in even $D=2d$ dimensions }

As known, the propagator of a relativistic spinning particle is the causal 
Green's function $S^{c}(x,y)$ of the Dirac equation.  In $D$ dimensions
the equation for this function has the form
\begin{equation}\label{b1} 
\left( \hat{{\cal P}}_\mu \gamma ^\mu - m\right)
S^{c}(x,y)=-\delta^D(x-y),
\end{equation}
where $\hat{{\cal P}}_{\mu}=i\partial _{\mu}-gA_{\mu}(x)$, 
$\mu=0,\ldots,D-1$, and
$\gamma^\mu$ are $\gamma$-matrices in $D$ dimensions,  $
\left[\gamma^\mu,\gamma^\nu\right]_+=2\eta^{\mu\nu}\;.$ 
Thus, in fact, we meet here the problem how to deal with an inverse
operator (to the Dirac one   $ \hat{{\cal P}}_\mu \gamma ^\mu -m$ )     
which has a complicated $\gamma$-matrix structure. As it is known
\cite{BW},  in even dimensions  a matrix
representation of the Clifford algebra with dimensionality 
${\rm dim}\,\gamma^{\mu}=2^{D/2}=2^d$ always exists. In other words 
${\gamma^{\mu}}$ are $2^d\times 2^d$ matrices. In such dimensions one
can introduce another matrix, $\gamma^{D+1}$, which anticommutes with
all $\gamma^{\mu}$ (analog of  $\gamma^5$ in four dimensions),
\begin{eqnarray}\label{b2}
&&\gamma^{D+1}=r\gamma^{0}\gamma^{1}\ldots \gamma^{D-1},\;
\; r=\left\{ \begin{array}{ll}
1,& \mbox{if $d$ is even} \\
i,& \mbox{if $d$ is odd}
\end{array}\right.\,, \\
&&[\gamma^{D+1},\gamma^{\mu}]_+=0,\;
\left(\gamma^{D+1}\right)^2=-1\,. \nonumber
\end{eqnarray}  
The existence of the matrix $\gamma^{D+1}$ in even dimensions  allows
one to pass to the Dirac   operator
which is  homogeneous  in $\gamma$-matrices. Indeed, let us
rewrite the equation (\ref{b1}) in terms of  the transformed by
$\gamma^{D+1}$ propagator  $\tilde{S}^{c}(x,y)$,
\begin{equation}\label{b3} 
\tilde{S}^{c}(x,y) = S^{c}(x,y)\gamma^{D+1},\;\;   
\left( \hat{{\cal P}}_\mu \tilde{\gamma}^\mu - m\gamma^{D+1} \right)
\tilde{S}^{c}(x,y)= \delta^{D}(x-y),
\end{equation}
where  $\tilde{\gamma}^
\mu = \gamma^{D+1}\gamma^\mu$. The matrices $\tilde{\gamma}^\mu$   
have the same 
commutation relations as initial ones $\gamma^\mu ,\; \;
\left[\tilde{\gamma}^\mu,\tilde{\gamma}^\nu\right]_+=2\eta^{\mu\nu}$,
and anticommute with the matrix $\gamma^{D+1}$. The set of $D+1$
$\gamma$-matrices $\tilde{\gamma}^\nu$ and $\gamma^{D+1}$ form a
representation of the Clifford algebra in odd $2d+1$
dimensions. Let us denote such matrices via $\Gamma^{n}$, 
\begin{eqnarray}\label{b4}
&&\Gamma^{n}=\left\{ \begin{array}{ll}
\tilde{\gamma}^{\mu},& n=\mu=0,\ldots,D-1 \\
\gamma^{D+1},& n=D
\end{array}\right.\;, \\
&&[\Gamma^k,\Gamma^n]_+ = 2\eta^{kn}, \; 
\; \eta_{kn}={\rm diag}(\underbrace{1,-1,\ldots,-1}_{D+1}),\;k,n=0,\ldots,D \,.  \nonumber 
\end{eqnarray}
In terms of these matrices the
equation (\ref{b3}) takes the form
\begin{equation}\label{b5} 
\hat{{\cal P}}_n \Gamma ^n  
\tilde{S}^{c}(x,y)= \delta^{D}(x-y),\;\;\hat{{\cal P}}_{\mu}=i\partial
_{\mu}-gA_{\mu}(x),\;\; \hat{\cal P}_D=-m\;.
\end{equation}
Now again, similar to (\ref{1b}), we present $\tilde{S}^{c}(x,y)$
as a matrix element of an operator $\hat{S}^{c}$ (in the
coordinate representation (\ref{2b})),
\begin{equation}\label{b6}
\tilde{S}^{c}_{ab}(x,y) 
= <x | \hat{S}^{c}_{ab} | y >,\;\; a,b=1,2,\ldots,2^d\,,
\end{equation}
where the spinor indices $a,b$ are written here explicitly for clarity  
and will be omitted hereafter. The equation (\ref{b5}) implies  
$ \hat{S}^{c} =\hat{F}^{-1},\;\;\hat{F}= \Pi_n\Gamma^n \, , $
where $\Pi_{\mu}$ are defined in (\ref{2b}), and
$\Pi_D=-m$. The operator $\hat{F}$ is
homogeneous in  $\gamma$-matrices, we can consider it as  a
pure Fermi one, if one reckons $\gamma$-matrices as Fermi-operators. 
Now, instead
of the Schwinger proper-time representation (\ref{3b}) which is convenient for
Bose-type operators  one can use a different representation by means of an
integral over the super-proper time 
($\lambda,\chi$)  of an exponential with an even exponent. Namely, one
can write\footnote{Here and in what follows
$\Pi^2=\Pi_{\mu}\Pi^{\mu}$ and so on.}
\begin{eqnarray}\label{b8}
&&\hat{S}^c = \frac{\hat{F}}{\hat{F}^2}=\int_0^\infty \, d\lambda 
\int  e^{i[\lambda(\hat{F}^2+i\epsilon)+ \chi\hat{F}]} d\,\chi\,, \\
&&\hat{F}^{2}=\Pi^2 - m^2-\frac{ig}{2}F_{\mu\nu}
\Gamma^\mu\Gamma^\nu\,, \nonumber
\end{eqnarray}
where $\lambda$ is an even variable and $\chi$ is an odd one, the 
latter anticommutes with $\hat{F}$ (with $\gamma$-matrices) by the
definition. Here and in what follows integrals 
over odd variables are understood as Berezin's integrals \cite{Ber2}. 
The representation (\ref{b8}) is 
an analog of the Schwinger proper-time representation for the inverse
operator  
convenient in the Fermi case.  Such a representation was introduced for the first time in
the paper \cite{FG}.    
Thus, the Green function (\ref{b6}) takes the form
\begin{eqnarray}\label{b10}
&&\tilde{S}^{c}=\tilde{S}^{c}(x_{out},x_{in}) =\int_0^\infty \, d\lambda 
\int \langle x_{\rm out} | e^{-i\hat{\cal H}(\lambda,\chi)}|x_{\rm in} 
\rangle d\chi\,,\\
&&\hat{{\cal H}}(\lambda,\chi)=\lambda \left(
m^2 - \Pi^2 +\frac{ig}{2}F_{\mu\nu}  \Gamma^\mu\Gamma^\nu\right) + 
\Pi_n\Gamma^n \,\chi\;.\nonumber
\end{eqnarray}
Now one can present the matrix element entering in the expression
(\ref{b10}) by means of a path integral. In spite of the fact that the operator
$\hat{{\cal H}}(\lambda,\chi)$ has  $\gamma$-matrix structure, it is 
possible to proceed as in scalar case. An analog of the formula
(\ref{6b}) has the form
\begin{eqnarray}\label{b11}
&& \tilde{S}^{c} =  \lim_{N\rightarrow \infty}\int_0^\infty d\, \lambda_0 
\int d\, \chi_0 \int_{-\infty}^{+\infty} d\, x_1 ... d\, x_{N-1}
d\, \lambda_1 ... d\, \lambda_N \int d\, \chi_1 ... d\, \chi_N \nonumber \\
&&\times \prod_{k=1}^{N}\langle x_k |
e^{-i\hat{{\cal H}}(\lambda_k,\chi_k)\Delta \tau} | x_{k-1} \rangle 
\delta(\lambda_k-\lambda_{k-1})\delta(\chi_k-\chi_{k-1})\; .
\end{eqnarray}
The matrix elements in (\ref{b11}) can be replaced by the expressions 
\begin{equation}\label{b13}
\int \frac{d\,p_k}{(2\pi)^D}\exp \left\{ i \left[
p_k\frac{x_k-x_{k-1}}{\Delta\tau} - {\cal H}(\lambda_k,\chi_k,
\overline{x}_k,p_k) \right]\Delta\tau\right\} \,,
\end{equation}
where ${\cal H}(\lambda,\chi,{x},{p})$ is the Weyl symbol of the 
operator $\hat{{\cal H}}(\lambda,\chi)$ in the sector of coordinates and 
momenta,
\[
{\cal H}(\lambda,\chi,x,p) =  \lambda \left(m^2-{\cal P}^2 +
\frac{ig}{2}F_{\mu\nu}  \Gamma^\mu\Gamma^\nu\right)
+ {\cal P}_n\Gamma^n \chi\;,
\]
and ${\cal P}_\mu= -p_\mu-gA_\mu(x),\; {\cal P}_D=-m\;. $
The multipliers (\ref{b13}) are noncommutative  due to the $\gamma$-matrix structure and are
situated in (\ref{b11}) so that the numbers $k$ increase 
from the right to the left. For the
two $\delta$-functions, accompanying each matrix element (\ref{b13}) in the
expression (\ref{b11}), we use the integral representations
$$
\delta(\lambda_k-\lambda_{k-1})\delta(\chi_k-\chi_{k-1}) = 
\frac{i}{2\pi} \int e^{i\left[ 
\pi_k\left(\lambda_k-\lambda_{k-1}\right)+
\nu_k\left(\chi_k   -\chi_{k-1}   \right)
\right]}d\, \pi_k d\, \nu_k,
$$
where $\nu_k$ are odd variables. Then we attribute formally the index $k$,  to 
$\gamma$-matrices, entering into  (\ref{b13}),  and
then we attribute to all quantities the ``time'' $\tau_k$, according
to the index $k$ they have, $\tau_k=k\Delta\tau $, so that $\tau \in [0,1]$.
Introducing the T-product which acts on $\gamma$-matrices it is possible 
to gather all the expressions, entering in  (\ref{b11}), 
in one exponent and deal then with the $\gamma$-matrices
like with odd  variables. Thus, we get for the right side of (\ref{b11})
\begin{eqnarray}\label{b14}
&&\tilde{S}^{c} = {\rm T}\int_0^\infty \, d\lambda_0 \int  
d\chi_{0}\int_{x_{in}}^{x_{out}}Dx \int Dp \int_{\lambda_0}D\lambda
\int_{\chi_0}D\chi\int D\pi \int D\nu \nonumber \\
&&\times\exp \left\{i\int_0^1 \left[ \lambda \left({\cal P}^2 - m^2 
-\frac{ig}{2}F_{\mu\nu}  \Gamma^\mu\Gamma^\nu
\right)+ \chi {\cal P}_n\Gamma^n  + 
p\dot{x} + \pi\dot{\lambda} + \nu\dot{\chi}\right]d\tau \right \} \,,  
\end{eqnarray}
where   $x(\tau)$,  $p(\tau)$,  $\lambda(\tau)$,  $\pi(\tau)$ ,
are even and $\chi(\tau), \;\nu(\tau)$ are odd trajectories, obeying the boundary conditions
 $\,\,\,x(0)=x_{\rm in}$, $\,\,\,x(1)=x_{\rm out}$, 
$\,\,\,\lambda (0) = \lambda_0$, 
$\,\,\,\chi(0) = \chi_0$. The operation of T-ordering  
acts on the $\gamma$-matrices which are supposed formally to
depend on time $\tau$. The expression (\ref{b14}) can be transformed then as follows:
\begin{eqnarray*}
&& \tilde{S}^{c} = \int_0^\infty \, d\lambda_0
\int d\chi_{0}\int_{x_{in}}^{x_{out}}Dx \int Dp \int_{\lambda_0}D\lambda
\int_{\chi_0}D\chi\int D\pi \int D\nu \exp \left\{i\int_0^1 \left[
\lambda\left( {\cal P}^2 -m^2 \right.\right. \right. \\
&&\left.\left.\left.-\frac{ig}{2}F_{\mu\nu} \frac{\delta_l}{\delta 
\rho_\mu}\frac{\delta_l}{\delta \rho_\nu}\right)   
   +\chi {\cal P}_n\frac{\delta_l}{\delta \rho_n} +\left. 
p\dot{x} + \pi\dot{\lambda} + \nu\dot{\chi}\right]d\tau\right\} 
{\rm T}\exp \int_0^1\rho_n(\tau)\Gamma^n d\tau \right|_{\rho=0},
\end{eqnarray*}
where  odd sources $\rho_n(\tau)$ are introduced. They  
anticommute with the $\gamma$-matrices by definition. One can present the 
quantity ${\rm T}\exp \int_0^1 \rho_n(\tau)\Gamma^n 
d\tau$  via a  path integral over odd trajectories \cite{FG,Vas}, 
\begin{eqnarray}\label{b15}
&&{\rm T}\exp \int_0^1\rho_n(\tau)\Gamma^n d\tau  = \exp\left(i\Gamma^n
\frac{\partial_l}{\partial\theta^n}   \right)
\int_{\psi(0)+\psi(1)=\theta}\exp \left[ \int_0^1 \left( 
\psi_n\dot{\psi}^n - 2i\rho_n\psi^n\right) d\tau \right.\nonumber \\ 
&&+ \left.\left.\psi_n(1)\psi^n(0)\right]{\cal D}\psi\right|_{\theta=0},\;\;
{\cal D}\psi=D\psi\left[\int_{\psi (0)+\psi (1)=0}D\psi \exp\left\{\int^{1}
_{0}\psi_{n}\dot{\psi}^{n}d\tau\right\}\right]^{-1} \; ,
\end{eqnarray}
where $\theta^n$ are odd variables, anticommuting with $\gamma$-matrices, and 
$\psi^{n}(\tau)$ are odd trajectories of integration, obeying the boundary 
conditions, which are pointed out below the signs of the integration. 
Using (\ref{b15}) we get the Hamiltonian path integral 
representation for the Green function in question:
\begin{eqnarray}\label{b15a}
&&\tilde{S}^{c} =\exp\left(i\Gamma^n
\frac{\partial_l}{\partial\theta^n} \right)\int_0^\infty \, d\lambda_0 
\int d\chi_{0}\int_{\lambda_{0}}D\lambda
\int_{\chi_{0}}D\chi \int_{x_{in}}^{x_{out}}Dx \int Dp \int D\pi \int
D\nu \nonumber \\ 
&&\times\int_{\psi(0)+\psi(1)=\theta} {\cal D}\psi \exp \left\{i\int_0^1 
\left[ \lambda\left({\cal P}^2 - m^2 +
2igF_{\mu\nu}\psi^\mu\psi^\nu\right)+2i{\cal P}_n
\psi^n \chi \right.\right. \nonumber \\ 
&&-i\psi_n\dot{\psi}^n + \left.\left.p\dot{x} + \pi \dot{\lambda} +
\nu \dot{\chi}
\right] d\tau +\left.\psi_n(1)\psi^n(0) \right\}\right|_{\theta=0}\,. 
\end{eqnarray}
Integrating over momenta, we get the path integral in the Lagrangian form,
\begin{eqnarray}\label{b16}
&&\tilde{S}^{c}=\exp\left(i\Gamma^{n}
\frac{\partial_{\ell}}{\partial \theta^{n}}\right)\int_{0}^{\infty}de_{0}
\int d\chi_{0}\int_{e_{0}} M(e)
De\int_{\chi_{0}}D\chi \int_{x_{in}}^{x_{out}}Dx \int D\pi \int 
D\nu  \nonumber \\  
&&\times \int_{\psi(0)+\psi(1)=\theta} {\cal D}\psi
\, \exp\left\{i\int_{0}^{1}\left[-\frac{\dot{x}^{2}}
{2e}-\frac{e}{2}m^{2}-g\dot{x}A+iegF_{\mu \nu}\psi^{\mu}\psi^{\nu} \right.\right. 
\nonumber \\
&&\left.\left. +i\left(\frac{\dot{x}_{\mu}\psi^{\mu}}{e}-m\psi^{D}\right)\chi
-i\psi_{n}\dot{\psi}^{n}+\pi \dot{e}+\nu \dot{\chi}\right]d\tau
+ \left.\psi_{n}(1)\psi^{n}(0)\right\}\right|_{\theta=0}\;,
\end{eqnarray}
\noindent where the measure $M(e)$ is defined by the eq. (\ref{12b}).

\subsection{Spinor case in odd $D=2d+1$ dimensions }

In odd dimensions  a possibility to construct the matrix
$\gamma^{D+1}$ (\ref{b2}) does not exist. Hence, the trick which was used to make the
Dirac operator homogeneous in $\gamma$-matrices does not work
here. Nevertheless, the problem of the path integral construction may
be solved in a different way. 

As it is known, in odd dimensions $D=2d+1$ there exist two exact non-equivalent
irreducible representations of the Clifford algebra with the
dimensionality $2^{[D/2]}=2^d$. Let us mark these representations by
the index $s=\pm $. Thus, we have two non-equivalent sets of
$\gamma$-matrices which we are going to denote as
$\Gamma_{(s)}^n,\;n=0,1,\ldots,2d$ (remark that now we use Latin indices
$n,k$, and so on, as Lorentz ones). Such matrices can be constructed,
e.g. from the corresponding matrices in $D=2d$ dimensions as follows:
\begin{eqnarray}\label{b17}
&&\Gamma^{n}_{(s)}=\left\{ \begin{array}{ll}
\gamma^{\mu},& n=\mu=0,\ldots,D-1 \\
s\gamma^{D+1},& n=D
\end{array}\right.\,, \\
&&[\Gamma_{(s)}^k,\Gamma_{(s)}^n]_+ = 2\eta^{kn}\,, \; 
 \eta_{kn}={\rm diag}\,(\underbrace{1,-1,\ldots,-1}_{D+1})\,,\;k,n=0,\ldots,D 
\;. \nonumber
\end{eqnarray}
In odd dimensions there exists also a duality relation which is important
for our purposes
\begin{equation}\label{b18}
\Gamma_{(s)}^n=\frac{sr}{(2d)!}\epsilon^{nk_1\ldots
k_{2d}}\Gamma_{(s)k_1}\ldots \Gamma_{(s)k_{2d}},\;\; 
r=\left\{ \begin{array}{ll}
1,& \mbox{if $d$ is even} \\
i,& \mbox{if $d$ is odd} 
\end{array}\right. \,.
\end{equation}  
Here $\epsilon^{nk\ldots l}$ is the Levi-Civita tensor density in $D$
dimensions.

The propagator $S^c(x,y)$ obeys the Dirac equation in the dimensions
under consideration  
\begin{equation}\label{b19} 
\left( \hat{{\cal P}}_n \Gamma_{(s)} ^n - m\right)
S^{c}(x,y)=-\delta^D(x-y),
\end{equation}
where $\hat{{\cal P}}_n=i\partial _n-gA_n(x)$. Thus, we get for the
operator $\hat{S}^c$ entering in (\ref{b6}),
$\hat{S}^c=-\hat{F}^{-1},\;\hat{F}=\Pi_n\Gamma_{(s)}^n-m\,,$
where all the $\Pi_n $ are defined by the equations (\ref{2b}). 
In the case under consideration it is convenient to present the
inverse operator in the following form
\begin{eqnarray}\label{b20}
&&\hat{S}^c=\frac{\hat{F}_{(+)}}{-\hat{F}_{(+)}\hat{F}}=
s\,\frac{\hat{A}}{\hat{B}}\;, \;\;\;\hat{F}_{(+)}=\Pi_n\Gamma_{(s)}^n+m\;, \\
&&\hat{A}= \frac{r}{(2d)!}\epsilon^{nk_1\ldots
k_{2d}}\Pi_n\Gamma_{(s)k_1}\ldots \Gamma_{(s)k_{2d}} +sm\,,\nonumber\\
&&\hat{B}=
m^2 - \Pi^2 +\frac{ig}{2}F_{kn}  
\Gamma_{(s)}^k\Gamma_{(s)}^n\,. \nonumber
\end{eqnarray}
The form of the operator $\hat{A}$ in (\ref{b20}) was obtained from the
operator $\hat{F}_{(+)}$ by means of the
duality relation (\ref{b18}). Now both
operators $\hat{A}$ and $\hat{B}$ are even in $\gamma$-matrices, 
so we can treat them as Bose-type operators. For their ratio
we are going to use a new kind of integral representation  which is
a combination of the Schwinger type  (\ref{3b}) representation  for
$\hat{B}^{-1}$ 
and additional representation of the operator $\hat{A}$ by means
of a Gaussian
integral over two Grassmannian variables $\chi_1$ and
$\chi_2$ with the involution property $ (\chi_1)^+=\chi_2$. Namely, one can write
\begin{equation}\label{b21}
\hat{S}^c=s\int_0^\infty d\lambda \int e^{-i[\lambda \hat{B}+\chi
\hat{A}]}d\chi\,,\;\;
\chi=\chi_1 \chi_2\,,\;\;d\chi=d\chi_1\,d\chi_2\,.  
\end{equation}
Thus, we get for the propagator 
\begin{eqnarray}\label{b22}
&&S^{c} =s\int_0^\infty \, d\lambda 
\int \langle x_{\rm out} | e^{-i\hat{\cal H}(\lambda,\chi)}|x_{\rm in} 
\rangle d\chi \;,\\
&&\hat{{\cal H}}(\lambda,\chi)=\lambda \left(
m^2 - \Pi^2 +\frac{ig}{2}F_{kn}  
\Gamma_{(s)}^k\Gamma_{(s)}^n\right) + 
\chi\left(\frac{r}{(2d)!}\epsilon^{nk_1\ldots
k_{2d}} \Pi_n \Gamma_{(s)k_1}\ldots \Gamma_{(s)k_{2d}} 
+sm \right)\;.\nonumber
\end{eqnarray}
Starting from this point one can proceed similarly to the even-dimensional
case to construct a path integral for the right side of
(\ref{b22}). The Hamiltonian form of such path integral is
\begin{eqnarray}\label{b23}
&&S^{c} =s\,\exp\left(i\Gamma_{(s)}^n
\frac{\partial_l}{\partial\theta^n} \right)\int_0^\infty \, d\lambda_0 
\int d\chi_{0}\int_{\lambda_{0}}D\lambda
\int_{\chi_{0}}D\chi \int_{x_{in}}^{x_{out}}Dx \int Dp \int D\pi \int
D\nu  \\ 
&&\times\int_{\psi(0)+\psi(1)=\theta} {\cal D}\psi \exp \left\{i\int_0^1 
\left[ \lambda\left({\cal P}^2 - m^2 +
2igF_{kn}\psi^k\psi^n\right)-\chi\left(sm
 \right.\right.\right. \nonumber \\ 
&&\left.+r\frac{(2i)^{2d}}{(2d)!} \epsilon^{nk_1\ldots
k_{2d}}{\cal P}_n\psi_{k_1}\ldots \psi_{k_{2d}}   \right)
-i\psi_n\dot{\psi}^n + \left.\left.p\dot{x} + \pi \dot{\lambda} +
\nu \dot{\chi}
\right] d\tau +\left.\psi_n(1)\psi^n(0)
\right\}\right|_{\theta=0}\,,\nonumber 
\end{eqnarray}
where   $x(\tau)$,  $p(\tau)$,  $\lambda(\tau)$,  $\pi(\tau)$ 
are even and $\psi(\tau),\;\chi_1(\tau),\,\chi_2(\tau),
\;\nu_1(\tau),\nu_2(\tau)$ are odd trajectories, obeying the boundary conditions
 $x(0)=x_{\rm in},\;x(1)=x_{\rm out},\;\lambda (0) =
\lambda_0,\;\chi(0) = \chi_0,\;\psi(0)+\psi(1)=\theta $, and the
notations are used
\[
\chi=\chi_1\chi_2,\;
\nu\dot{\chi}=\nu_1\dot{\chi}_1+\nu_2\dot{\chi}_2,\;
d\chi=d\chi_1\,d\chi_2,\;D\chi=D\chi_1\,D\chi_2,\;D\nu=D\nu_1\,D\nu_2\,.
\]
Integrating over momenta, we get a path integral in the Lagrangian form,
\begin{eqnarray}\label{b24}
&&S^{c}=\frac{s}{2}\,\exp\left(i\Gamma_{(s)}^{n}
\frac{\partial_{\ell}}{\partial \theta^{n}}\right)\int_{0}^{\infty}de_{0}
\int d\chi_{0}\int_{e_{0}} M(e)
De\int_{\chi_{0}}D\chi \int_{x_{in}}^{x_{out}}Dx \int D\pi \int 
D\nu  \\  
&&\times \int_{\psi(0)+\psi(1)=\theta} {\cal D}\psi
\, \exp\left\{i\int_{0}^{1}\left[-\frac{\dot{x}^{2}}
{2e}-\frac{e}{2}m^{2}-g\dot{x}_n A^n+iegF_{kn}\psi^{k}\psi^{n} \right.\right. 
\nonumber \\
&&\left.\left. -\chi\left(sm+\frac{r}{e}\frac{(2i)^{2d}}{(2d)!}\epsilon^{nk_1\ldots
k_{2d}}\dot{x}_n\psi_{k_1}\ldots \psi_{k_{2d}}\right)
-i\psi_{n}\dot{\psi}^{n}+\pi \dot{e}+\nu \dot{\chi}\right]d\tau
+ \left.\psi_{n}(1)\psi^{n}(0) \right \}\right|_{\theta=0}\;,\nonumber
\end{eqnarray}
\noindent where the measure $M(e)$ is defined by the eq. (\ref{12b})
and $e(0)=e_0$.  

One can also get a different form of the path integral for the Dirac
propagator in odd dimensions. To this end, instead of (\ref{b20}), one
has to write 
\begin{eqnarray}\label{b24a}
&&\hat{S}^c=\frac{\hat{F}}{-\hat{F}\hat{F}}=
s\,\frac{\hat{A}}{\hat{B}}\,,\\
&&\hat{A}= \frac{r}{(2d)!}\epsilon^{nk_1\ldots
k_{2d}}\Pi_n\Gamma_{(s)k_1}\ldots \Gamma_{(s)k_{2d}} -sm\,,\nonumber\\
&&\hat{B}=
m^2 - \Pi^2 +\frac{ig}{2}F_{kn}  
\Gamma_{(s)}^k\Gamma_{(s)}^n-\frac{2srm}{(2d)!}\epsilon^{nk_1\ldots
k_{2d}}\Pi_n\Gamma_{(s)k_1}\ldots \Gamma_{(s)k_{2d}}  \,, \nonumber
\end{eqnarray}
and then proceed as before. Thus, we get one more  form of the Lagrangian
path integral 
\begin{eqnarray}\label{b24b}
&&S^{c}=\frac{s}{2}\,\exp\left(i\Gamma_{(s)}^{n}
\frac{\partial_{\ell}}{\partial \theta^{n}}\right)\int_{0}^{\infty}de_{0}
\int d\chi_{0}\int_{e_{0}} M(e)
De\int_{\chi_{0}}D\chi \int_{x_{in}}^{x_{out}}Dx \int D\pi \int 
D\nu  \\  
&&\times \int_{\psi(0)+\psi(1)=\theta} {\cal D}\psi
\, \exp\left\{i\int_{0}^{1}\left[-\frac{\dot{x}^{2}}
{2e}-\frac{e}{2}m^{2}-g\dot{x}_n A^n(x)+iegF_{kn}(x)\psi^{k}\psi^{n}+sm\chi \right.\right. 
\nonumber \\
&&\left.\left. -\left(\frac{\chi}{e}-sm\right)\frac{r(2i)^{2d}}{(2d)!}\epsilon^{nk_1\ldots
k_{2d}}\dot{x}_n\psi_{k_1}\ldots \psi_{k_{2d}}
-i\psi_{n}\dot{\psi}^{n}+\pi \dot{e}+\nu \dot{\chi}\right]d\tau
+ \left.\psi_{n}(1)\psi^{n}(0) \right \}\right|_{\theta=0}\;.\nonumber
\end{eqnarray}

\section{Pseudoclassical description of spinning particles in
arbitrary dimensions}

The path integral representations for particles propagators  have also
 an important heuristic
value. They give  a possibility to guess the form of actions to
 describe the particles classically or pseudoclassically if we believe
that such representations should have the form $\int \exp
(iS)\,D\varphi$. Here $\varphi$ is a set of variables and $S$ a
classical action. Indeed, let
us take the simplest example of the scalar particle. Here the path integral
representation of the propagator has the form (\ref{11b}). The
exponent in the integrand of the path integral can be treated as a
Lagrangian action of the relativistic spinless particle. This exponent
consists of two parts. The first one 
\begin{equation}\label{c1}
S=-\int_{0}^{1}\left [
-\frac{\dot{x}^2}{2e}-e\frac{m^2}{2}
-g\hat{x}_{\mu}A^{\mu} \right]d\,\tau
\end{equation}
is  well-known gauge-invariant (reparametrization-invariant)
action of the relativistic spinless particle. The corresponding gauge
transformations read $\delta x = \dot{x}\xi \,,\; 
\delta e = \frac{d}{d\tau}(e\xi)$.
The second term in the exponent can be treated as a gauge-fixing term 
which corresponds to the gauge condition $\dot{e}=0$. Quantization of
the action (\ref{c1}) leads to the corresponding quantum theory of a
scalar particle \cite{GT1}  which is equivalent to  one-particle
sector of the scalar quantum field theory. Thus, we have a closed
circle: propagator (which is a representative of the one-particle
sector of the scalar quantum field theory) - path integral for it -
classical action of a point-like particle - quantization - one-particle
sector of the scalar quantum field theory. 

Let us now turn to the case of spinning particle, which is certainly
of the main interest for us,  using the experience
in the simple spinless case. Namely, looking on  the path integrals (\ref{b16}) or
(\ref{b24})  we can guess the form of the actions for spinning
particles  in even and odd dimensions. Hence, the exponent in the
integrand of the right side of (\ref{b16}) can be treated as a
pseudoclassical action of the spinning particle  in even
dimensions. Separating the gauge-fixing terms and boundary terms, we
get a gauge-invariant pseudoclassical action
\begin{eqnarray}\label{c3}
&&S=\int_{0}^{1}\left[-\frac{z^{2}}
{2e}-\frac{e}{2}m^{2}-g\dot{x}_{\mu}A^{\mu}+iegF_{\mu
\nu}\psi^{\mu}\psi^{\nu}
 -im\psi^{D}\chi
-i\psi_{n}\dot{\psi}^{n}\right]d\tau\;, \nonumber \\
&&z^{\mu}=\dot{x}^{\mu}-i\psi^{\mu}\chi\;.
\end{eqnarray}
There are two type of gauge transformations in the 
theory with the action (\ref{c3}): reparametrizations,
\begin{equation}\label{c4}
\delta x^{\mu} = \dot{x}^{\mu}\,\xi \,,\;\; 
\delta e = \frac{d}{d\tau}(e\xi) \,,\;\; 
\delta \psi^n = \dot{\psi}^n\xi  \,,\;\;
\delta \chi = \frac{d}{d\tau}(\chi\xi)\, ,
\end{equation} 
and supertransformations,
\begin{equation}\label{c5}
\delta x^\mu = i\psi^\mu\epsilon \,,\;
\delta e = i\chi \epsilon \,,\;
\delta \chi = \dot{\epsilon} \,,\; 
\delta\psi^\mu = \frac{z^\mu}{2e}\epsilon\,, \; 
\delta\psi^D = \frac{m}{2}\epsilon\,,
\end{equation}
where $\xi$ is even and $\epsilon$ is odd $\tau$-dependent parameters.

The action (\ref{c3}) is a trivial generalization of the BM action to $D$ dimensions. The
quantization of the action (\ref{c3}) in even dimensions can be done \cite{GGr} completely
similar to the four-dimensional case \cite{GT1}. It reproduces the
quantum theory of spinning particle (in particular, the Dirac
equation), which is  equivalent to the one-particle sector
of the spinor field quantum theory, with the propagator $S^c$
(\ref{b1}).

In odd dimensions, the path integral  (\ref{b24}) prompt us the
following pseudoclassical action to describe spinning particles in
such dimensions:
\begin{eqnarray}\label{c6}
&&S=\int_{0}^{1}\left[-\frac{z^{2}}
{2e}-\frac{e}{2}m^{2}-g\dot{x}_nA^n+iegF_{kn}\psi^{k}\psi^{n}
 -\kappa m \chi
-i\psi_{n}\dot{\psi}^{n}\right]d\tau=\int_{0}^{1}\, L d\tau\;, \nonumber \\
&&z^{n}=\dot{x}^{n}+r\frac{(2i)^{2d}}{(2d)!}\epsilon^{nk_1\ldots
k_{2d}}\psi_{k_1}\ldots \psi_{k_{2d}} \chi\;.
\end{eqnarray}
We   suppose that $\kappa$ is an even  constant,  which
will be discussed below   and   
$\chi=\chi_1\chi_2$, with $\chi_1$ and $\chi_2$ being Grassmannian
variables, obeying the involution properties:
$\chi_1^{+}=\chi_2$. Interpreting the variable $\chi$ in such a way,
we can discover that the action (\ref{c6}) is gauge-invariant
(reparametrization- and supergauge-invariant). The corresponding gauge
transformations have the form: reparametrizations 
\begin{eqnarray}\label{c7}
&&\delta x^{n} = \dot{x}^{n}\,\xi \,,\;\; 
\delta e = \frac{d}{d\tau}(e\xi) \,,\;\; 
\delta \psi^n = \dot{\psi}^n\xi  \,, \nonumber \\
&&\delta \chi_1 = \dot{\chi}_1\xi +\frac{1}{2}\chi_1\dot{\xi}\, ,\;\;
\delta \chi_2 = \dot{\chi}_2\xi +\frac{1}{2}\chi_2\dot{\xi}\,,
\end{eqnarray} 
and two sets of nonlocal (in time) supertransformations,
\begin{eqnarray}\label{c8}
&&\delta x^n = i \epsilon^{nk_1\ldots
k_{2d}}\psi_{k_1}\ldots \psi_{k_{2d}} U\,,\;\;
\delta\psi^n =-\frac{d}{e}\epsilon^{nk_1k_2\ldots
k_{2d}}z_{k_1}\psi_{k_2}\ldots \psi_{k_{2d}} U\,,\nonumber \\
&&\delta e =0 \,,\;
\delta \chi_1=\theta_1,\;\;\delta \chi_2=\theta_2,\;\;
U=\frac{ir(2i)^{2d}}{(2d)!}\int_{0}^{\tau}[\chi_1\theta_2-\chi_2\theta_1]d\tau\,,
\end{eqnarray}
where $\xi$ is even and $\theta_{1,2}$ are odd $\tau$-dependent
parameters.  

As was already mentioned in the Introduction, formerly there were
proposed three different kinds of action to describe spinning particles
in odd dimensions \cite{Pl1,Pl2,GGT1}. It has been shown that an
adequate quantum theory arises in course of those actions
quantization 
(at least the Dirac equation appears). However it was not  demonstrated how one
can construct the  path integral for the propagator by means of those
actions (path integral quantization was not done). The action of the
paper \cite{Pl2} is close enough to the action (\ref{c6}), however,
contains additional dynamical Grassmann variables, and $\chi$ is not
interpreted as a composit bifermionoc-type variable. We already
have proved that our new action (\ref{c6}) allows one to write the 
corresponding
path integral and now we are going to check that the direct (operator)
quantization leads to the corresponding quantum theory. To this end,
as usual, we need  to analyse the Hamiltonian structure of the
theory with the action (\ref{c6}).

Introducing the canonical momenta
\begin{eqnarray}\label{c9}
&&p_n = \frac{\partial L}{\partial\dot{x}^n} = -\frac{z_n}{e} -
gA_n(x)\, , \;\;
P_n =  \frac{\partial_r L}{\partial\dot{\psi}^n} = -i\psi_n\,,
\nonumber \\
&&P_e = \frac{\partial L}{\partial\dot{e}} = 0,\; 
P_{\chi_{1,2}} = \frac{\partial_r L}{\partial\dot{\chi}_{1,2}} = 0\,,
\end{eqnarray}
one can see  that there exist primary 
constraints  $\Phi^{(1)}=0$,
\begin{equation}\label{c10}
\Phi_{1,2}^{(1)}= P_{\chi_{1,2}}\,, \; \; \Phi_3^{(1)}= P_e\,, \;\;
\Phi_{4n}^{(1)}= P_n +i\psi_n  \;.
\end{equation}
We construct the total  Hamiltonian according to the standard procedure
\cite{Dir,GT2} (we use the notations of the book \cite{GT2}), and get 
$H^{(1)}=H + \lambda \Phi^{(1)}$, with
$$
H = -\frac{e}{2}\left({\cal P}^2 - m^2 +
2igeF_{kn}\psi^k\psi^n\right)+\chi\left(r\frac{(2i)^{2d}}{(2d)!} \epsilon^{nk_1\ldots
k_{2d}}{\cal P}_n\psi_{k_1}\ldots \psi_{k_{2d}} +\kappa m
   \right)\,,
$$
where ${\cal P}=-p_{n}-gA_{n}(x)$. From the consistency conditions  (Dirac procedure) 
we find a set of independent secondary constraints $\Phi^{(2)} = 0$,
\begin{equation}\label{c11}
\Phi_1^{(2)} ={\cal P}^2 - m^2 +
2igeF_{kn}\psi^k\psi^n \,   ,  \;\;
\Phi_2^{(2)}=r\frac{(2i)^{2d}}{(2d)!} \epsilon^{nk_1\ldots
k_{2d}}{\cal P}_n\psi_{k_1}\ldots \psi_{k_{2d}}+\kappa m
 \,.
\end{equation}
One can go over from the initial set of constraints 
$\left(\Phi^{(1)},\Phi^{(2)}\right)$ to  the equivalent one
$ \left(\Phi^{(1)},\tilde{\Phi}^{(2)}\right)$, where
$\tilde{\Phi}^{(2)}=\Phi^{(2)}(\psi\rightarrow \psi +
\frac{i}{2}\Phi_4^{(1)}) $. The new set of 
constraints can be explicitly divided in a set of  
first-class constraints, which is $\left(\Phi^{(1)}_{1,2,3}, \;\tilde{\Phi}^{(2)} \right)$ and 
in a set of second-class constraints, which is $\Phi^{(1)}_{4n}$.

Let us consider the Dirac quantization, where the second-class
constraints define the Dirac brackets and therefore the commutation relations.
The first-class constraints, being applied to the state vectors,
define physical states. Thus, we get for essential operators and nonzero
commutation relations: 
\begin{equation}\label{c12}
[\hat{x}^\mu,\hat{p}_\nu]=i\{x^\mu,p_\nu\}_{D(\Phi^{(1)}_4)}=
i\delta^\mu_\nu\;, \;\;
[\hat{\psi}^k,\hat{\psi}^n]_+=i\{\psi^k,\psi^n\}_{D(\Phi^{(1)}_4)}=
-\frac{1}{2}\eta^{kn}\;. 
\end{equation}
According to the scheme of quantization selected, operators of the
second-class constraints are identically zero, whereas the operators of the 
first-class constraints have to annul  physical state
vectors. Taking that into account, one may 
construct a realization of the commutation relations
(\ref{c12}) in a Hilbert space ${\cal R}$ whose elements ${\bf f}\in 
{\cal R}$ are $2^d$-component columns dependent only on $x$, such that
\begin{equation}\label{c13}
\hat{x}^n=x^n{\bf I}\,,\;\; \hat{p}_n=-i\partial_n{\bf I}\,,
\;\; \hat{\psi}^n=\frac{i}{2}\Gamma_{(s)}^n\,, 
\end{equation}
where  ${\bf I}$ is
$2^d\times 2^d$ unit matrix, and  $\Gamma_{(s)}^n$,  are $\gamma$-matrices
in $D=2d+1$ dimensions, see (\ref{b17}). Besides of that, we have the
following equations for the physical state vectors 
\begin{equation}\label{c14}
\hat{\Phi}_1^{(2)}{\bf f}(x)=0\,,\;\; \hat{\Phi}_2^{(2)}{\bf
f}(x)=0\,,
\end{equation}
where $\hat{\Phi}^{(2)}$ are operators, which correspond to the constraints
(\ref{c11}). Taken into account the duality relation (\ref{b18}), one
can write the second equation (\ref{c14}) in the form
\begin{equation}\label{c15}
\left( \hat{{\cal P}}_n \Gamma_{(s)} ^n +s\kappa m\right){\bf f}(x)=0\,,
\end{equation}
where $\hat{{\cal P}}_n=i\partial _n-gA_n(x)$. Thus, we get the 
Dirac equation in course of the operator quantization if we put
$\kappa=-s$ in course of the quantization. By this choice of $\kappa$ the first equation
(\ref{c14}) is simply a consequence of the Dirac equation. It is the
squared Dirac equation.

\section{Consideration for  spinning particles with anomalous magnetic
moment}

One can generalize the construction of path integrals to the case of
the particles with an anomalous magnetic moment (AMM). In four
dimensions the corresponding pseudoclassical models were discussed in
\cite{Bar,Zhe,GiS,Gri}. The path integral representation for the
propagator was constructed in \cite{GiS2}. We may use our approach
starting with the generalized  by Pauli \cite{pauli}  Dirac equation and considering it in
$D$ dimensions:
\begin{equation}\label{Ap1}
 \left[ \hat{{\cal P}}_\nu \gamma^\nu -\left( m +
\frac{\mu}{2}\sigma^{\alpha\beta}
F_{\alpha\beta}\right) \right]\Psi(x)= 0\,.
\end{equation} 
Here $\sigma^{\alpha\beta}=
\frac{i}{2}[\gamma^\alpha,\gamma^\beta]_-\,$, and $\mu$  stands for
AMM. As before we have to consider two cases of even and
odd dimensions separately. In even dimension $D=2d$ one has to modify  the equation
(\ref{b3}) to the following form 
\begin{equation}\label{Ap2}    
\left[ \hat{{\cal P}}_\mu \tilde{\gamma}^\mu - \gamma^{D+1}\left(m
+i\frac{\mu}{2}F_{\alpha\beta}\tilde{\gamma}^{\alpha}\tilde{\gamma}^{\beta}
 \right)\right]
\tilde{S}^{c}(x,y)= \delta^{D}(x-y),
\end{equation}
Proceeding
as before, we get the following generalization of the path integral
representation (\ref{b16})
\begin{eqnarray}\label{Ap3}
&&\tilde{S}^{c}=\exp\left(i\Gamma^{n}
\frac{\partial_{\ell}}{\partial \theta^{n}}\right)\int_{0}^{\infty}de_{0}
\int d\chi_{0}\int_{e_{0}} M(e)
De\int_{\chi_{0}}D\chi \int_{x_{in}}^{x_{out}}Dx \int D\pi \int 
D\nu  \nonumber \\  
&&\times \int_{\psi(0)+\psi(1)=\theta} {\cal D}\psi
\, \exp\left\{i\int_{0}^{1}\left[-\frac{\dot{x}^{2}}
{2e}-\frac{e}{2}M^{2}+iegF_{\mu
\nu}\psi^{\mu}\psi^{\nu}-\dot{x}^{\alpha}\left(gA_{\alpha}+4i\mu\psi^DF_{\alpha
\beta}\psi^{\beta}\right)
\right.\right.  \nonumber \\
&&\left.\left.
+i\left(\frac{\dot{x}_{\mu}\psi^{\mu}}{e}-
M^*\psi^{D}\right)\chi
-i\psi_{n}\dot{\psi}^{n}+\pi \dot{e}+\nu \dot{\chi}\right]d\tau
+ \left.\psi_{n}(1)\psi^{n}(0)\right\}\right|_{\theta=0}\;,
\end{eqnarray}
where $M=m-2i\mu F_{\alpha\beta}\psi^{\alpha}\psi^{\beta}$. Thus, a
pseudoclassical action for spinning particle with AMM has the form in
even dimensions
\begin{eqnarray}\label{Ap4}
&&S=\int_{0}^{1}\left[-\frac{z^{2}}
{2e}-\frac{e}{2}M^{2}- \dot{x}^{\alpha}\left(gA_{\alpha}
+4i\mu\psi^DF_{\alpha
\beta}\psi^{\beta}\right) \right. \nonumber \\
&& \left. +iegF_{\mu
\nu}\psi^{\mu}\psi^{\nu}
 -iM^*\psi^{D}\chi
-i\psi_{n}\dot{\psi}^{n}\right]d\tau\;, \;\;z^{\mu}=\dot{x}^{\mu}-i\psi^{\mu}\chi\;.
\end{eqnarray}

In odd dimensions $D=2d+1$ we have to modify the equation (\ref{b19}),
introducing Pauli term, 
\begin{equation}\label{Ap5} 
\left[ \hat{{\cal P}}_n \Gamma_{(s)} ^n - 
\left(m+i\frac{\mu}{2}F_{kn}\Gamma_{(s)} ^k\Gamma_{(s)} ^n\right )\right]
S^{c}(x,y)=-\delta^D(x-y)\,.
\end{equation}
Thus, we get for the
operator $\hat{S}^c$ entering in (\ref{b6}),
\[
\hat{S}^c=-\hat{F}^{-1},\;\hat{F}=\Pi_n\Gamma_{(s)}^n-
\left(m+i\frac{\mu}{2}F_{kn}\Gamma_{(s)} ^k\Gamma_{(s)} ^n \right)\,,
\]
where all  $\Pi_n $ are defined by  eq. (\ref{2b}). Using
 duality relation (\ref{b18}) we can  present $\hat{S}^c$ in the following form
\begin{eqnarray}\label{Ap6}
&&\hat{S}^c=\frac{\hat{F}_{(+)}}{-\hat{F}_{(+)}\hat{F}}=
s\,\frac{\hat{A}}{\hat{B}}\;, \;\;\;\hat{F}_{(+)}=\Pi_n\Gamma_{(s)}^n+
\left(m+i\frac{\mu}{2}F_{kn}\Gamma_{(s)} ^k\Gamma_{(s)} ^n \right)
 \;, \\
&&\hat{A}= \frac{r}{(2d)!}\epsilon^{nk_1\ldots
k_{2d}}\Pi_n\Gamma_{(s)k_1}\ldots \Gamma_{(s)k_{2d}} +
s\left(m+i\frac{\mu}{2}F_{kn}\Gamma_{(s)} ^k\Gamma_{(s)} ^n \right)\,,\nonumber\\
&&\hat{B}=
m^2 - \Pi^2 +i\left(m\mu+\frac{g}{2}\right)F_{kn}  
\Gamma_{(s)}^k\Gamma_{(s)}^n
-\frac{\mu^2}{4}\left(F_{kn}\Gamma_{(s)} ^k\Gamma_{(s)} ^n\right)^2
\nonumber \\
&&-i\frac{s\mu r}{(2d)!}\epsilon^{kk_1\ldots
k_{2d}}[F_{kn},\Pi^n]_{+}\Gamma_{(s)k_1}\ldots \Gamma_{(s)k_{2d}}
\,. \nonumber
\end{eqnarray}
Then the representation (\ref{b22}) takes place, where $\hat{{\cal
H}}(\lambda,\chi)=\lambda \hat{B} + 
\chi\hat{A}$, with $\hat{A}$ and $\hat{B}$ defined by
(\ref{Ap6}). Proceeding similarly to the case without AMM we get the
Hamiltonian form of the path integral for $S^c$,
\begin{eqnarray}\label{Ap7}
&&S^{c} =s\,\exp\left(i\Gamma_{(s)}^n
\frac{\partial_l}{\partial\theta^n} \right)\int_0^\infty \, d\lambda_0 
\int d\chi_{0}\int_{\lambda_{0}}D\lambda
\int_{\chi_{0}}D\chi \int_{x_{in}}^{x_{out}}Dx \int Dp \int D\pi \int
D\nu  \\ 
&&\times\int_{\psi(0)+\psi(1)=\theta} {\cal D}\psi \exp \left\{i\int_0^1 
\left[ \lambda\left({\cal P}^2 - M^2 +
2igF_{kn}\psi^k\psi^n \right.\right.\right. \nonumber \\
&&\left. + \frac{s\mu r(2i)^D }{(2d)!}\epsilon^{kk_1\ldots
k_{2d}}F_{kn}{\cal P}^n\psi_{k_1}\ldots \psi_{k_{2d}}\right)-\chi\left(sM
 +r\frac{(2i)^{2d}}{(2d)!} \epsilon^{nk_1\ldots
k_{2d}}{\cal P}_n\psi_{k_1}\ldots \psi_{k_{2d}} \right)  \nonumber \\ 
&&\left.\left.-i\psi_n\dot{\psi}^n + 
p\dot{x} + \pi \dot{\lambda} +
\nu \dot{\chi}
\right] d\tau +\left.\psi_n(1)\psi^n(0)
\right\}\right|_{\theta=0}\,.\nonumber 
\end{eqnarray}
Here  the following notations are used
\[
\chi=\chi_1\chi_2,\;
\nu\dot{\chi}=\nu_1\dot{\chi}_1+\nu_2\dot{\chi}_2,\;
d\chi=d\chi_1\,d\chi_2,\;D\chi=D\chi_1\,D\chi_2,\;D\nu=D\nu_1\,D\nu_2\,.
\]

Integrating over momenta, we get a path integral in the Lagrangian form,
\begin{eqnarray}\label{Ap8}
&&S^{c}=\frac{s}{2}\,\exp\left(i\Gamma_{(s)}^{n}
\frac{\partial_{\ell}}{\partial \theta^{n}}\right)\int_{0}^{\infty}de_{0}
\int d\chi_{0}\int_{e_{0}} M(e)
De\int_{\chi_{0}}D\chi \int_{x_{in}}^{x_{out}}Dx \int D\pi \int 
D\nu  \\  
&&\times \int_{\psi(0)+\psi(1)=\theta} {\cal D}\psi
\, \exp\left\{i\int_{0}^{1}\left[-\frac{\dot{x}^{2}}
{2e}-\frac{e}{2}M^{2}+iegF_{kn}\psi^{k}\psi^{n} -\dot{x}^n\left(gA_n \right.\right.\right. 
\nonumber \\
&&\left.\left.\left. +\frac{s\mu r(2i)^D }{2(2d)!}F_{nk}\epsilon^{kk_1\ldots
k_{2d}}\psi_{k_1}\ldots \psi_{k_{2d}}\right)
 -\chi\left(sM+\frac{r}{e}\frac{(2i)^{2d}}{(2d)!}\epsilon^{nk_1\ldots
k_{2d}}\dot{x}_n\psi_{k_1}\ldots \psi_{k_{2d}}\right) \right.\right. \nonumber \\
&&\left.\left.-i\psi_{n}\dot{\psi}^{n}+\pi \dot{e}+\nu \dot{\chi}\right]d\tau
+ \left.\psi_{n}(1)\psi^{n}(0) \right\}\right|_{\theta=0}\;,\nonumber
\end{eqnarray}
Thus, a
pseudoclassical action for spinning particle with AMM has the
following  form in odd dimensions
\begin{eqnarray}\label{Ap9}
&&S=\int_{0}^{1}\left[-\frac{z^{2}}
{2e}-\frac{e}{2}M^{2} -\dot{x}^n\left(
gA_n  +\frac{s\mu r(2i)^D }{2(2d)!}F_{nk}\epsilon^{kk_1\ldots
k_{2d}}\psi_{k_1}\ldots \psi_{k_{2d}}\right)\right. \nonumber \\
&&\left. +iegF_{kn}\psi^{k}\psi^{n} -s M \chi
-i\psi_{n}\dot{\psi}^{n}\right]d\tau\,, \;\;
z^{n}=\dot{x}^{n}+r\frac{(2i)^{2d}}{(2d)!}\epsilon^{nk_1\ldots
k_{2d}}\psi_{k_1}\ldots \psi_{k_{2d}} \chi\;.
\end{eqnarray}

Quantizing both actions (\ref{Ap4}) and (\ref{Ap9}) one reproduces the
Dirac Pauli equation (\ref{Ap1}) similarly to the case without AMM.

\section{Discussion}

One of the new features in the 
pseudoclassical model in odd dimensions (\ref{c6})  consists in a new 
interpretation of the even variable $\chi$ as a composite bifermionic
type variable. One ought to say that only from the point of view of the
operator quantization, one may treat $\chi$ as an unique bosonic
variable. Indeed, one can believe that $\chi$ is simply an ordinary
Lagrange multiplier as it occurred always before. Doing the Dirac
procedure, bearing in mind this interpretation, one gets finally the same first class
constraint (\ref{c11}) and the same Dirac brackets. Thus, the result of
the Dirac quantization will be the same. However, as was demonstrated
above, the  new interpretation is necessary for the path
integral construction, or for path integral quantization. Besides of
that it provides the desirable
supersymmetry of the pseudoclassical action. Indeed, treating $\chi$
as an unique even variable, we have the following  symmetries of the
action (\ref{c6}): reparametrizations
\begin{equation}\label{sym}
\delta x^{n} = \dot{x}^{n}\,\xi \,,\;\; 
\delta e = \frac{d}{d\tau}(e\xi) \,,\;\; 
\delta \psi^n = \dot{\psi}^n\xi,\;\delta \chi=\frac{d}{d\tau}(\chi\xi)\,,
\end{equation} 
and gauge transformations,
\begin{eqnarray}\label{tr}
&&\delta x^n = i \epsilon^{nk_1\ldots
k_{2d}}\psi_{k_1}\ldots \psi_{k_{2d}} \beta\,,\;\;
\delta\psi^n =-\frac{d}{e}\epsilon^{nk_1k_2\ldots
k_{2d}}z_{k_1}\psi_{k_2}\ldots \psi_{k_{2d}} \beta\,,\nonumber \\
&&\delta e =0 \,,\;
\delta \chi=-\frac{i(2d)!}{r(2i)^{2d}}\dot{\beta}\,,
\end{eqnarray}
where both  $\xi$  and $\beta $ are even $\tau$-dependent
parameters. Thus, in this case no supersymmetry is present. One ought
to say that the presence of the nonlocal supersymmetry (\ref{c8}) in the
model (\ref{c6}) is a characteristic feature of the pseudoclassical
theory only, since it was proved in \cite{GT2} that for singular
theories with bosonic variables any gauge transformations are local in
time. Treating  $\chi$ as a composite bifermionic
type variable,  we meet also  for the
first time  a situation when the action and Hamiltonian are
quadratic in  Lagrange multipliers.

Another question is how to  interpret the  constant $\kappa$ in the action
(\ref{c6}). This question is directly related with the well-known
problem of classical inconsistency of some kind of constraints in
pseudoclassical mechanics. Indeed, if one treats $\kappa$ as an
ordinary complex parameter, then  from the classical point of view
 the constraint equation 
$\Phi_2^{(2)}=0$ is inconsistent. 
Such a difficulty appears not for the first time in the
pseudoclassical mechanics, (see for example \cite{HPPT}). Here the
following point of view is possible. One can believe that in classical
theory $\kappa$ is an even, bifermionic type element of the Berezin
algebra, $\kappa^2=0$. Then the above-mentioned constraint equation
appears to be consistent in the classical theory. At the same time, as
was pointed out in \cite{GGT2}, one has to admit a possibility to
change the nature of the parameters  in course of transition from the 
pseudoclassical mechanics to
the quantum theory (why we  admit such a possibility for the dynamical
variables?). Namely, in quantum theory the parameter $\kappa$ appears
to be  a real number, whose possible values are defined by the
quantum dynamics. For example, the path integral quantization of the
action (\ref{c6}) demands $\kappa\rightarrow s$, whereas the operator
quantizations demands $\kappa\rightarrow -s$, where $s=\pm 1$ defined
 an irreducible representation of the Clifford algebra, see
(\ref{b17}). To get the same quantization for $\kappa$ both in path
integral quantization and operator quantization (at given and fixed
choice of the irreducible representations for $\gamma$ matrices) one
has to consider another action, which can be extracted from the
alternative path integral representation  (\ref{b24b}). Such an action
has the form 
\begin{eqnarray}\label{alt}
&&S=\int_{0}^{1}\left[-\frac{z^{2}}
{2e}-\frac{e}{2}m^{2}-g\dot{x}_nA^n+iegF_{kn}\psi^{k}\psi^{n}
 +\kappa m \chi
-i\psi_{n}\dot{\psi}^{n}\right]d\tau=\int_{0}^{1}\, L d\tau\;, \nonumber \\
&&z^{n}=\dot{x}^{n}+r\frac{(2i)^{2d}}{(2d)!}\epsilon^{nk_1\ldots
k_{2d}}\psi_{k_1}\ldots \psi_{k_{2d}} (\chi-\kappa e m)\;,
\end{eqnarray}
with the same  variables and parameters.

One may  also note that path integral representations for
particles propagators have not only a pure theoretical interest. It 
makes possible to  calculate effectively these propagators in various
configurations of external fields (see for example \cite{All3}). Such
propagators are necessary composite parts for calculations  in
quantum field theory with non-perturbative backgrounds
\cite{All4}. That is why the representations for the propagators
obtained can be useful   in quantum field theory in   higher
dimensions, which attract attention already for a long time since the Kaluza-Klein ideas.

The presented path integral representations  may be useful in 
the so called spin factor problem, which was opened first by  Polyakov
\cite{Pol}. He assumed that the propagator 
of a free Dirac electron in $D=3$ Euclidean space-time can 
be presented by means of a bosonic path integral similar to the 
scalar particle case, modified by a
so called spin factor. This idea was developed by several authors, see
for example \cite{All5}, in particular to derive  
the spin factor for spinning particles interacting with external
fields. Surprisingly, it was shown in  \cite{Gsh} that all the
Grassmannian integrations in the representation (\ref{b24})  of Dirac propagator
in an arbitrary external field in four dimensions can be  done, so that an expression 
for the spin factor was derived as a
given functional of the bosonic trajectory. Having such an expression
for the spin factor one can use it to calculate the propagator in some
particular cases of external fields \cite{GZC}. This way of calculation
automatically provides the explicit $\gamma$-matrix structure of the
propagators, that facilitate a lot  concrete calculations with the
propagators. The new path
integral representation  obtained by us in the present
paper allows one to get the expression for the spin factor in the same
manner already in arbitrary dimensions. The corresponding details will be
published soon.

{\bf Acknowledgements}

The author thanks Brazilian foundations  CNPq  for support.

\end{document}